\documentclass{emulateapj}

\usepackage{color}

\usepackage{soul}   %% use \st{}
\newcommand{\kimmag}{-1.5}
\newcommand{\kimdist}{105}
\newcommand{\kimmet}{$-1.0$}

\begin{document}
\title{DISCOVERY OF A FAINT OUTER HALO MILKY WAY STAR CLUSTER IN THE SOUTHERN SKY}

\author{Dongwon Kim} 
\author{Helmut Jerjen} 
\author{Antonino P. Milone} 
\author{Dougal Mackey} 
\author{Gary S. Da Costa} 
\affil{Research School of Astronomy and Astrophysics, The Australian National University, Mt Stromlo Observatory, via Cotter Rd, Weston, ACT 2611, Australia}

\email{dongwon.kim@anu.edu.au}

\begin{abstract}
We report the discovery of a new, low luminosity star cluster in the outer halo of the Milky Way. High quality $gr$ photometry is presented, from which a color-magnitude diagram is constructed, and estimates of age, [Fe/H], [$\alpha$/Fe], and distance are derived. The star cluster, which we designate as Kim\,2, lies at a heliocentric distance of $\sim105$\,kpc. With a half-light radius of $\sim12.8$\,pc and ellipticity of $\epsilon\sim0.12$, it shares the properties of  outer halo GCs, except for the higher metallicity ([Fe/H]$\sim-1.0$) and lower luminosity ($M_{V}\sim-1.5)$. These parameters are similar to those for the globular cluster AM\,4, that is considered to be associated with the Sagittarius dwarf spheroidal galaxy.
We find evidence of dynamical mass segregation and the presence of extra-tidal stars that suggests Kim\,2 is most likely a star cluster. Spectroscopic observations for radial-velocity membership and chemical abundance measurements are needed to further understand the nature of the object.

\end{abstract}


\section{Introduction}
Globular clusters in the outer halo of the Milky Way (MW) hold important clues to the formation and structure of their host galaxy. Most of these rare distant globular clusters exhibit anomalously red horizontal branch morphology at given metal abundance~\citep{Lee1994}, and belong to the so-called ``young halo" system~\citep{Zinn1993a}. Young halo objects are hypothesized to have formed in external dwarf galaxies that were accreted into the Galactic potential well and disrupted by the Galactic tidal force~\citep{Searle1978}. This scenario has received considerable support by observational results from the MW and M31~\citep{DaCosta1995,Marin-Franch2009,Mackey2004,Mackey2010}. Indeed, the young halo clusters resemble the globular clusters located in dwarf galaxies associated with the Milky Way in terms of horizontal branch type~\citep{Zinn1993b, Smith1998,Johnson1999, Harbeck2001} and other properties such as luminosity, age, and chemical abundance~\citep{DaCosta2003}.
   
Despite the significant contribution of modern imaging surveys like the Sloan Digital Sky Survey~\citep{Ahn2014} to the discoveries of new Milky Way satellite galaxies~\citep[e.g][]{Willman2005,Belokurov2007,Irwin2007,Walsh2007} and extended substructures\citep[e.g][]{Newberg2003,Grillmair2009}, only a small number of star clusters have been discovered~\citep{Koposov2007,Belokurov2010,Balbinot2013,Kim1}, and these are typically located in the inner halo of the Milky Way. A new distant MW halo object at 145\,kpc, by the name of PSO\,J174.0675-10.8774, or Crater, was recently discovered simultaneously in two independent surveys~\citep{Laevens2014,Belokurov2014}. Although this  stellar system shares the structural properties of globular clusters in the outer halo of the Galaxy, confirming its true nature still requires additional investigation. Other than PSO\,J174.0675-10.8774, only six known Milky Way GCs are located at Galactocentric distances larger than 50\,kpc, namely AM\,1, 
Eridanus, NGC\,2419, Palomar\,3, 4, and 14 (see Table \ref{tab:outerGCs}). The Hubble Space Telescope Advanced Camera for Survey photometry of the Galactic GCs~\citep{Sarajedini2007,Dotter2011} has confirmed that all of the outer halo GCs except for NGC\,2419 have a red horizontal branch and young ages relative to the inner halo GCs~\citep{Dotter2010}.

 \begin{deluxetable*}{lrrrrrrrl}
 \tablecolumns{9} 
\tablewidth{0pt}
\tablecaption{Properties of the seven most distant Galactic globular clusters known.}
\tablehead{
\colhead{} &
\colhead{} &
\colhead{} &
\colhead{} &
\colhead{} &
\colhead{} &
\colhead{} &
\colhead{PSO\,J174.0675} &
\colhead{} \\
\colhead{Parameter} & 
\colhead{AM\,1} &
\colhead{Pal 3} &
\colhead{Pal 4} &
\colhead{Pal 14} &
\colhead{Eridanus} &
\colhead{NGC2419} &
\colhead{-10.8774\tablenotemark{a}} &
\colhead{Unit}}
\startdata
$\alpha_{J2000}$ 	& 03 55 02.3	& 10 05 31.9	& 11 29 16.8 	& 16 11 00.6 	& 04 24 44.5  	& 07 38 08.4	& 11 36 16.2&h:m:s \\
$\delta_{J2000}$ 	& $-$49 36 55 	& $+$00 04 18 	& $+$28 58 25	& $+$14 57 28 	& $-$21 11 13 	& $+$38 52 57 & $-$10 52 39&$^\circ:\arcmin:\arcsec$ \\
$l$ & 258.34 &240.15&  202.31 & 28.74 & 218.10 & 180.37 &$274.8$ & deg\\
$b$ &  $-48.47$ & $+41.86$ & $+71.80$ & $+42.19$& $-41.33$ & $+25.24$& $+47.8$ & deg\\
$D_\odot$ & 123.3  & 92.5 & 108.7& 76.5 &90.1& 82.6&145& kpc \\
$D_{gc}$ & 124.6  &  95.7 & 111.2 &   71.6 &95.0& 89.9&145& kpc \\
{[Fe/H]} 	& $-1.70$	&$-1.63$	& $-1.41$ 	& $-1.62$	& $-1.43$	& $-2.15$	& $-1.9$&  dex \\
$r_{h}$(Plummer) & 15.2 & 18.0& 16.6 &28.0 & 12.4& 22.1& 22& pc \\
$M_{tot,V}$ & $-4.73$ &$-5.69$& $-6.01$ & $-4.80$& $-5.13$& $-9.42$&$-4.3$& mag 
\enddata

\label{tab:outerGCs}
\tablecomments{From~\citet[and 2010 edition]{Harris1996}, combined with \cite{Laevens2014} for PSO\,J174.0675-10.8774.}
\tablenotetext{a}{PSO\,J174.0675-10.8774 is not yet unambiguously confirmed as a globular cluster.}
\end{deluxetable*}

In this paper, we report the discovery of a distant globular cluster in the constellation of Indus. This object was first detected in our on-going southern sky blind survey with the Dark Energy Camera (DECam) at the 4\,m Blanco Telescope at Cerro Tololo Inter-American Observatory (CTIO) and confirmed with deep GMOS-S images at the 8.1\,m Gemini South telescope on Cerro Pach\~on, Chile (Section 2 \& 3). The new star cluster, which we designate as Kim\,2, is at a distance $D_\odot\sim\kimdist$\,kpc and has a low luminosity of only $M_{V}\sim\kimmag$\,mag and a metallicity of [Fe/H]$\approx$\kimmet, slightly higher than the other young halo clusters (Section 4). In section 5 we discuss the implication 
of these properties, present evidence for mass segregation in the cluster and discuss its possible origin. 

\begin{deluxetable}{ccccc}
\tablewidth{0pt}
\tablecaption{GMOS Observing Log for the Images used in the Analysis}
\tablehead{
\colhead{Filter} &
\colhead{UT Date} &
\colhead{Exposure} &
\colhead{Seeing} &
\colhead{Airmass}}
\startdata
$g$ & Sep 20 2014 & $9\times292$s & $0\farcs6$ - $0\farcs9$ & 1.08 - 1.12 \\
$r$ & Oct 29 2014 & $9\times292$s & $0\farcs8$ - $0\farcs9$ & 1.23 - 1.42 
\enddata
\label{tab:Log}
\end{deluxetable}

\section{Discovery}
As part of the Stromlo Milky Way Satellite Survey\footnote{http://www.mso.anu.edu.au/$\sim$jerjen/SMS\_Survey.html} we collected 
imaging data for $\sim$ 500 sqr deg with the DECam at the 4\,m Blanco telescope at CTIO over three photometric nights from 17th to 19th July 2014. DECam is an array of sixty-two 2k$\times$4k CCD detectors with a 2.2 deg$^2$ field of view and a pixel scale of $0\farcs27$(unbinned). We obtained a series of $3 \times 40$\,s dithered exposures in the $g$ and $r$ band under photometric conditions. The average seeing was $1\farcs0$ for both filters each night. The stacked images were reduced via the DECam community pipeline~\citep{DECamCP2014}. We used WeightWatcher~\citep{WeightWatcher} for weight map combination and SExtractor~\citep{SExtractor} for source detection and photometry. Sources were morphologically classified as stellar or non-stellar objects. For the photometric calibration, we regularly observed Stripe 82{\footnote{ http://cas.sdss.org/stripe82/en/}}  of the Sloan Digital Sky Survey throughout the three nights with 50\,s single exposures in each band. To determine zero points and color terms, we matched our instrumental
magnitudes with the Stripe 82 stellar catalogue to a depth of $\sim$ 23\,mag and fit the following equations: 

\begin{equation}
g = g_{instr} + zp_{g} + c_{g} (g_{instr} - r_{instr}) - k_{g}X
\end{equation}

\begin{equation}
r = r_{instr} + zp_{r} + c_{r} (g_{instr}- r_{instr}) - k_{r}X
\end{equation}

where $zp_{g}$ and $zp_{r}$ are the zero points, $c_{g}$ and $c_{r}$ are the respective color terms, $k_{g}$ and $k_{r}$ are the first order extinctions, and $X$ is the mean airmass. 

In the Stripe 82 images we observed right after the Kim\,2 field, we found 399 stars with $19<r<23$ and $0.0<g-r<2.0$ in the identical CCD chip where the cluster was detected. We restricted the calibration to stars fainter than $r=19$\,mag to avoid the saturation limit of our DECam data. 
We determined the zero points, color terms and associated uncertainties by bootstrapping with replacements performed 1000 times and using a linear least-squares 
fit with  3-sigma clipping rejection. Uncertainties in the zero points were measured 0.013\,mag in $g$ and 0.010 in $r$, whereas uncertainties in the 
color terms are 0.011 and 0.009, respectively. The most recent extinction values $k_g$ and $k_r$ for CTIO were obtained from the Dark Energy Survey team.
We calibrated our DECam photometry of the Kim\,2 field using these coefficients and corrected for exposure time differences.

We employ the same detection algorithm as described in \cite{Kim1} to search the photometry catalog for stellar overdensities. In essence, we apply a photometric filter in color-magnitude space adopting isochrone masks based on the Dartmouth stellar evolution models~\citep{Dartmouth} to enhance the presence of old and metal-poor stellar populations relative to 
the Milky Way foreground stars. We then bin the R.A., decl. positions of the stars and convolve the 2-D histogram with a Gaussian kernel. The statistical significance of potential overdensities is measured by comparing their signal to noise ratios (S/Ns) on the density map to those of random clustering in the residual Galactic foreground. This process is repeated for different bin sizes and Gaussian kernels by shifting the isochrone masks over a range of distance moduli $(m-M)$ from 16 to 22 magnitudes. We detected the new stellar overdensity with a significance of $\sim8\,\sigma$ relative to the Poisson noise of the Galactic foreground stars. This object that we chose to call Kim\,2 was found at 21h08m49.97s, $-$51d09m48.6s(J2000) in the constellation of Indus.

\begin{figure}
\begin{center}
\includegraphics[scale=0.3]{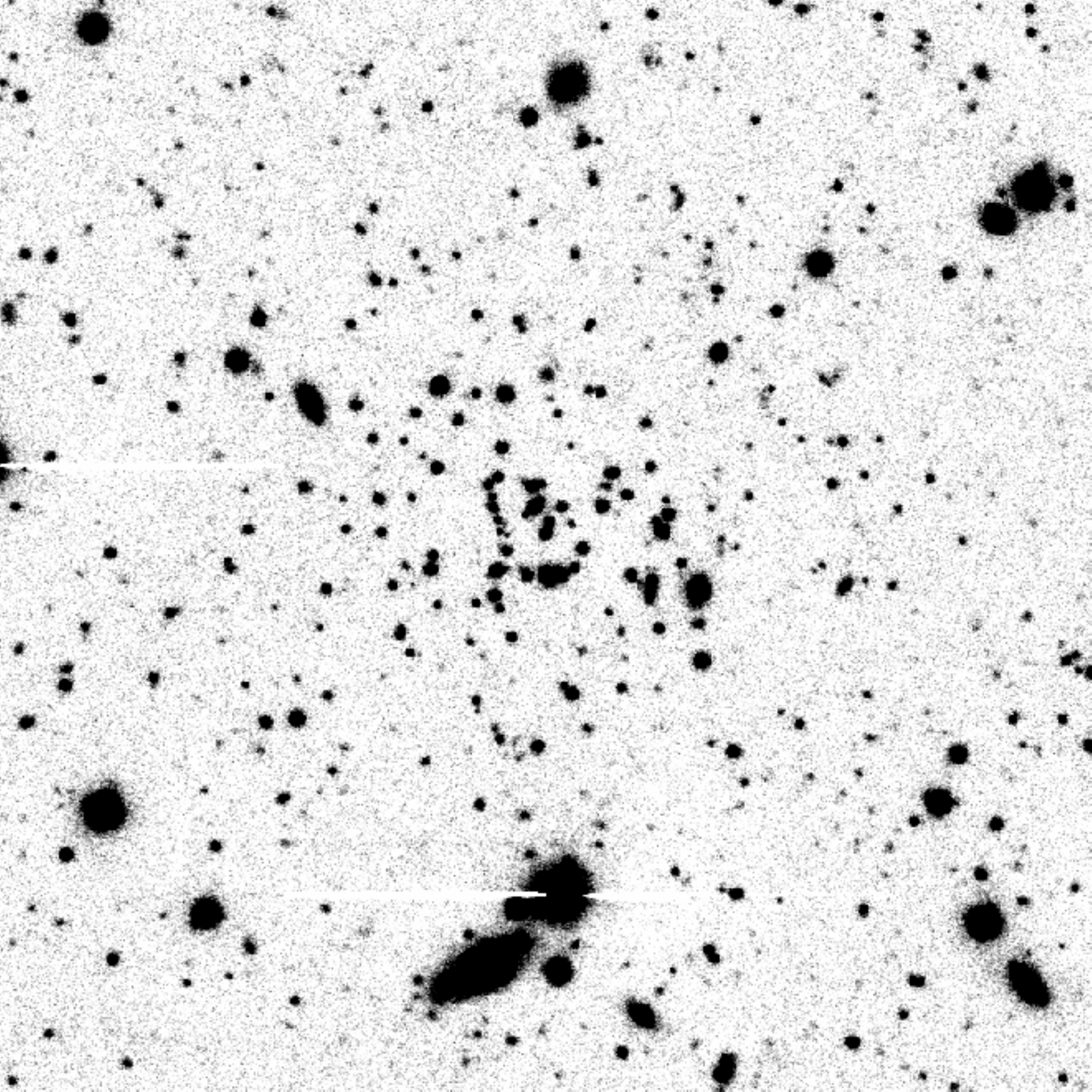}
\end{center}
\caption{$4\times 4$\,arcmin$^2$ GMOS cutout g-band image of Kim\,2. The cluster is located at the centre of the image. North is up, east is to the left.\label{fig:FitsImage}}
\end{figure}

\section{Follow-up Observations and Data Reduction}
To investigate the nature of Kim\,2, deep follow-up observations were obtained with the Gemini Multi-Object Spectrograph in imaging mode at the 8.1\,m Gemini South telescope through director's time (GS-2014B-DD-3) on Sep 20, 21, 22, 30 and Oct 29. Since June 2014, GMOS-S is equipped with a new array of three $2048\times4176$\, pixel$^2$ Hamamatsu CCDs with a $5\farcm5 \times 5\farcm5$ field of view and a pixel scale of $0\farcs08$ (unbinned). To reduce readout time, we employed $2\times2$ binning, resulting in a plate scale of $0\farcs16$ pixel$^{-1}$. A series of $17\times292$\,s dithered exposures in $g'$ and $19\times292$ in $r'$ band were observed. These $g'$ and $r'$ filters are similar, but not identical, to the $g$ and $r$ filters used by the SDSS.  We chose the nine best images in each band for our photometric analysis. A summary of the selected observations is presented in Table~\ref{tab:Log}. 

We employed the latest Gemini IRAF package V1.13 (commissioning release)\footnote{http://www.gemini.edu/node/12227} for data reduction. We applied bias and flat-field images provided by the Gemini science archive for standard GMOS baseline calibration to each exposure using the GIREDUCE task. The three CCD frames of each reduced image were then mosaicked into a single frame using the GMOSAIC task. Figure~\ref{fig:FitsImage} shows a cut out at the centre of a deep $g'$ band image, formed by combining the nine individual mosaicked frames of the passband using the IMCOADD task, in which Kim\,2 is visible as a concentration of faint stars.

The photometry of the reduced GMOS images was carried out using the software package {\it kitchen\_sync}  presented in~\cite{Anderson2008} and modified to work with GMOS-S data. It exploits two distinct methods to measure bright and faint stars. Astrometric and photometric measurements of bright stars have been performed in each mosaicked image, independently, by using appropriate point-spread function (PSF) model, and later combined. To derive the PSF models, we adapted to our data the software as described in~\cite[][see also~\citealt{Bellini2010}]{Anderson2006}. Briefly, we used the most isolated, bright and non-saturated stars in each image to determine a grid of four empirical PSFs. To account for the  spatial variation of the PSF across the field of view, we assumed that to each pixel of the image corresponds a PSF that is a bi-linear weighted interpolation of the closest four PSFs of the grid. 

Furthermore, the flux and position can also be determined by fitting for each star simultaneously all the pixels in all the exposures. This approach works better for very faint stars, which can not be robustly measured in every individual exposure.  We refer the reader to the papers by~\cite{Anderson2006} and \cite{Anderson2008} for further details.

We then conducted the photometric calibration using 65 stars with $22<r<25$ and $0.0<g-r<2.0$ we found in the field of view of GMOS.
Comparing their instrumental magnitudes to the calibrated magnitudes of our DECam photometry in Section 2, we derived a calibration equation composed of a photometric zero point and a color term from bootstrapping the data 1000 times and performing a least-square fit with 3-sigma clipping rejection. Uncertainties in the zero points are 0.022\,mag in the $g$ band and 0.023\,mag in $r$. Uncertainties in the color terms are 0.020 and 0.019, respectively.

\begin{figure}
\begin{center}
\includegraphics[scale=0.4]{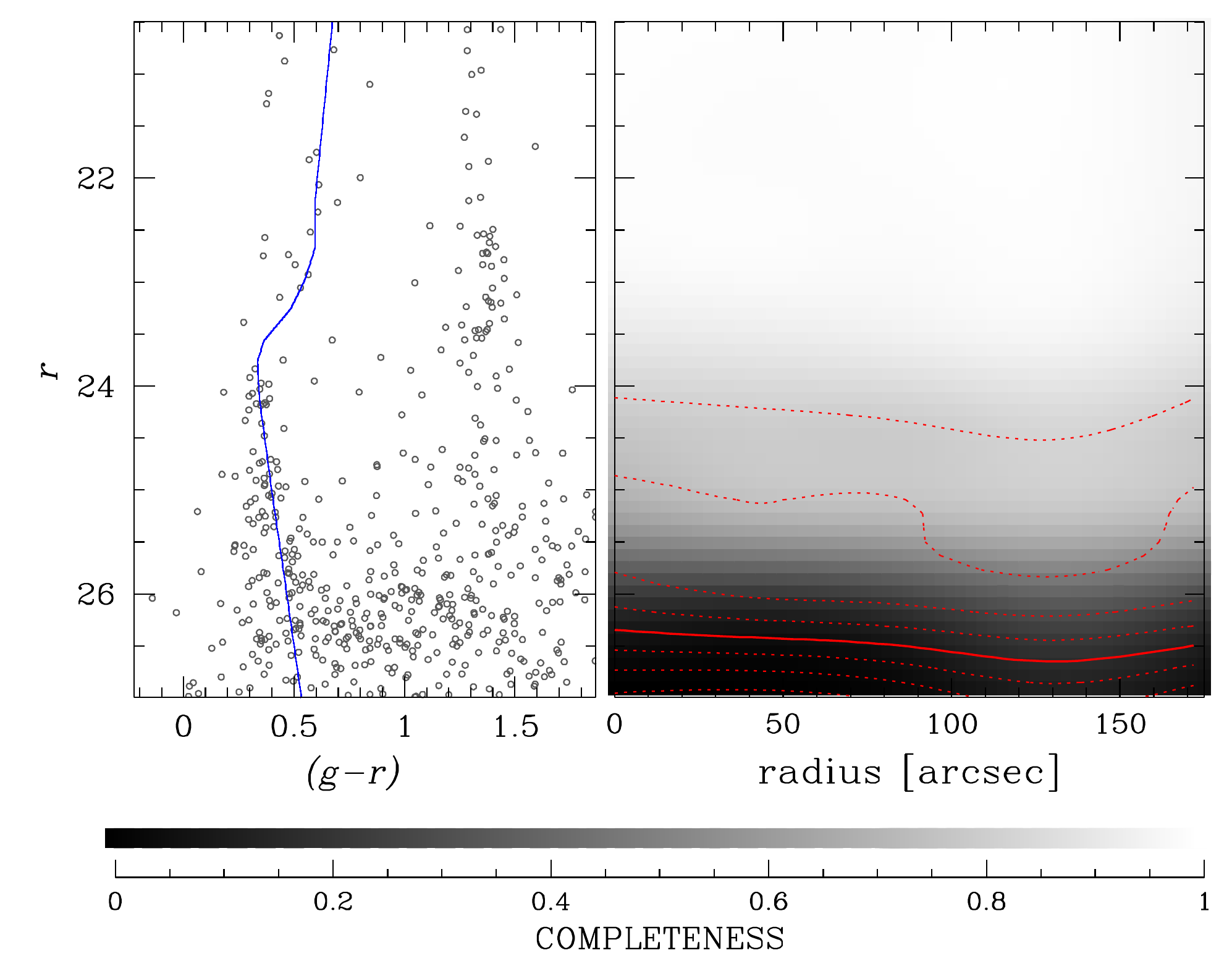}
\end{center}
\caption{Left panel: CMD in instrumental magnitudes. Overplotted on the CMD is the fiducial line along which artificial stars are added. Right panel: Completeness contours in the radial distance versus $r$ magnitude plane. The contour lines correspond to the completeness levels of $90\% - 20\%$. The solid contour line marks $50\%$ level. \label{fig:Completeness}}
\end{figure}

We performed artificial star tests to determine the completeness level of our photometry. To do this we used the recipe and the software described in detail by~\cite{Anderson2008}.  Briefly, we first generated an input list of artificial stars and placed along the fiducial line of the MS and the RGB of Kim\,2, which we have derived by hand. The list includes the coordinates of the stars in the reference frame and the magnitudes in g and r bands. Artificial stars have been placed in each image according to the overall cluster distribution as in~\cite{Milone2009}.

For each star in the input list, the software by~\cite{Anderson2008} adds, in each image, a star with appropriate flux and position and measures it by using the same procedure as for real stars. An artificial star is considered to be detected when the input and the output position differ by less then 0.5 pixel and the input and the output flux by less than 0.75 mag.

The software provides for artificial stars the same diagnostics of the photometric quality as for real star. We applied the same procedure used for real stars to select a sub-sample of stars with small astrometric errors, and well fitted by the PSF. Figure~\ref{fig:Completeness} shows the recovery rate of the input stars as a function of the stellar magnitude and the radial distance from the cluster center.

To address the effect of crowding, we measured the completeness not only as a function of the stellar magnitude 
but also the distance from the cluster center. For the latter, we divided the GMOS field into five concentric annuli, in each of which we measured the 
completeness in seven magnitude bins, in the interval $-14<g,r_{instr}<-5$. Interpolating the recovery rate of the input stars at each of these $7\times5$ grid points allows us to estimate the completeness of any star at any position within the cluster as shown in Figure~\ref{fig:Completeness}.

\section{Candidate Properties}

\subsection{Color-Magnitude Diagram}

\begin{figure*}
\begin{center}
\includegraphics[scale=0.85]{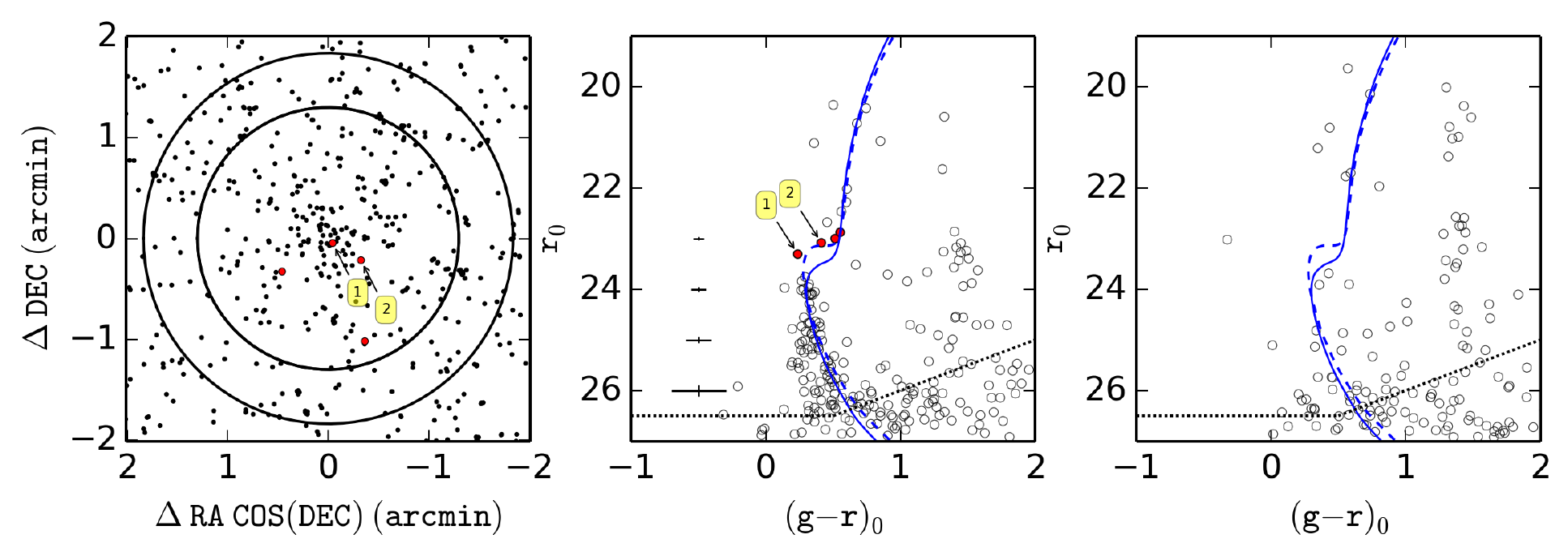}
\end{center}
\caption{GMOS view of Kim\,2. \emph{Left panel}: Distribution of all objects classified as stars in $2\arcmin\times2\arcmin$ field centred on the cluster. The circles mark a radius of $1\farcm3$\,($\sim3r_{h}$) and $1\farcm8$. \emph{Middle panel}: CMD of all stars within the inner-most circle marked on the left panel, dominated by the members of the star cluster. \emph{Right panel}: Comparison CMD of all stars between the inner and the outer circles, showing the foreground stars. The dotted lines mark the 50\% completeness level of our photometry. The two best-fitting Dartmouth isochrones with 11.5\,Gyr, [Fe/H]$=-1.0$, [$\alpha$/H]$=+0.2$ (solid line), and with 8.0\,Gyr, [Fe/H]$=-0.9$, [$\alpha$/H]$=+0.4$ (dashed line) are overplotted at a distance of 105 kpc and 98\,kpc, respectively. \label{fig:CMD}}
\end{figure*}

The left panel of Figure~\ref{fig:CMD} shows the RA-DEC distribution of all objects classified as point source by our GMOS photometry centred on Kim\,2. The middle panel of Figure~\ref{fig:CMD} shows the extinction-corrected CMD of all stars located within $1\farcm3$\,($\sim3r_{h}$) from the overdensity center. All magnitudes are individually corrected for Galactic reddening by the \cite{Schlegel1998} maps and the extinction coefficients of \cite{Schlafly2011}. In Table~\ref{tab:GMOSdata}, we present our GMOS photometry of all stars brighter than the 50\% completeness level, the dotted line in the middle panel of Figure~\ref{fig:CMD}. For comparison, the right panel shows the CMD of stars in an equal area between the radii $1\farcm3$ and $1\farcm8$, the majority of which are expected to be MW field stars.

The subgiant branch and the red giant branch (RGB) of this loose and faint cluster is almost absent, and no hints of an horizontal branch or red giant clump are visible. 
The main sequence (MS) however is well defined down to $r_0\approx26.5$, below which our photometry is affected by incompleteness. There are four possible subgiant branch and MS turn-off stars (red dots in Figure~\ref{fig:CMD}) consistent with the location of a main-sequence that runs from $r_{0}\sim23.5$\,mag down to $26.5$\,mag. 
The stars labelled \#1 and \#2 have small angular distances from the nominal cluster center (see left panel of Figure~\ref{fig:CMD}). This supports the idea that they are cluster members.
If true, we would observe a lack of stars between star \#1 and the brightest main sequence stars. Such a gap in the luminosity function is 
uncommon but not unheard of, for example in Segue\,3 \citep[see Fig.2 in ][]{Fadely2011}. Overplotted on our CMD are two theoretical isochrones from the Dartmouth 
data base. They will be discussed in the next section.

\begin{deluxetable*}{ccccccccccc}
\tablecaption{GMOS PHOTOMETRY FOR STARS WITHIN $3\,r_{h}$ FROM THE CENTER OF KIM\,2}
\tablewidth{0pt}
\tablehead{
\colhead{} &
\colhead{} &
\colhead{Radial} &
\colhead{} &
\colhead{} &
\colhead{} &
\colhead{} &
\colhead{} &
\colhead{Radial} &
\colhead{} &
\colhead{} \\
\colhead{$\alpha$ (J2000)} & 
\colhead{$\delta$ (J2000)} &
\colhead{Distance} &
\colhead{$r$} &
\colhead{$(g-r)$} &
\colhead{} &
\colhead{$\alpha$ (J2000)} & 
\colhead{$\delta$ (J2000)} &
\colhead{Distance} &
\colhead{$r$} &
\colhead{$(g-r)$} \\
\colhead{(h$\,$ $\,$ m$\,$ $\,$s)} &
\colhead{($^\circ\,$ $\,'\,$ $\,''$)} &
\colhead{($\arcmin$)} &
\colhead{(mag)} &
\colhead{(mag)} &
\colhead{} &
\colhead{(h$\,$ $\,$ m$\,$ $\,$s)} &
\colhead{($^\circ\,$ $\,'\,$ $\,''$)} &
\colhead{($\arcmin$)} &
\colhead{(mag)} &
\colhead{(mag)}} 
\startdata
21:08:49.81&-51:09:46.66&0.040&24.16&0.37&&21:08:46.11&-51:09:52.80&0.608&24.86&1.18\\
21:08:50.04&-51:09:51.57&0.051&24.42&0.46&&21:08:53.28&-51:10:08.69&0.618&23.31&1.52\\
21:08:49.78&-51:09:51.74&0.060&21.70&1.34&&21:08:51.43&-51:09:13.99&0.620&25.80&0.36\\
21:08:49.68&-51:09:51.33&0.064&23.38&0.27&&21:08:47.67&-51:09:18.26&0.621&26.35&0.52\\
21:08:50.42&-51:09:51.05&0.082&23.83&0.32&&21:08:46.14&-51:09:35.97&0.636&25.28&0.29\\
21:08:49.46&-51:09:50.63&0.087&24.96&0.37&&21:08:45.87&-51:09:39.90&0.658&25.83&0.55\\
21:08:49.50&-51:09:45.45&0.090&26.27&0.52&&21:08:46.96&-51:09:20.73&0.662&23.46&1.54\\
21:08:49.36&-51:09:48.45&0.095&24.18&0.38&&21:08:53.62&-51:09:26.57&0.681&21.15&0.88\\
21:08:49.61&-51:09:43.60&0.100&24.71&0.40&&21:08:47.79&-51:09:13.09&0.683&26.09&0.31\\
21:08:50.46&-51:09:43.76&0.111&25.22&0.42&&21:08:48.20&-51:10:26.09&0.683&26.14&0.54\\
21:08:50.45&-51:09:53.91&0.117&23.97&0.35&&21:08:53.07&-51:10:17.52&0.685&25.48&0.50\\
21:08:49.99&-51:09:41.23&0.123&24.12&0.39&&21:08:50.20&-51:09:07.17&0.691&25.85&0.51\\
21:08:50.48&-51:09:42.44&0.130&25.52&0.61&&21:08:50.17&-51:10:30.25&0.695&20.67&1.36\\
21:08:50.39&-51:09:55.39&0.131&25.50&0.49&&21:08:54.46&-51:09:55.98&0.715&25.08&0.32\\
21:08:50.52&-51:09:41.00&0.154&24.94&0.37&&21:08:49.22&-51:09:05.58&0.727&25.98&0.54\\
21:08:49.12&-51:09:43.51&0.158&23.91&0.30&&21:08:54.43&-51:09:31.98&0.752&25.52&0.23\\
21:08:50.41&-51:09:40.04&0.158&24.81&0.43&&21:08:50.76&-51:10:33.49&0.758&25.60&0.47\\
21:08:50.23&-51:09:58.76&0.174&24.90&0.33&&21:08:46.43&-51:09:15.23&0.785&21.19&0.39\\
21:08:51.21&-51:09:49.31&0.195&25.09&0.45&&21:08:49.35&-51:09:00.85&0.802&24.07&0.31\\
21:08:48.76&-51:09:51.53&0.196&24.36&0.35&&21:08:49.65&-51:10:37.71&0.820&20.50&0.77\\
21:08:51.24&-51:09:50.87&0.203&24.09&0.29&&21:08:55.20&-51:09:48.95&0.820&23.64&1.43\\
21:08:50.36&-51:09:36.70&0.208&24.17&0.33&&21:08:50.02&-51:08:58.14&0.841&22.75&0.48\\
21:08:48.81&-51:09:42.16&0.211&24.74&0.34&&21:08:53.89&-51:10:23.19&0.843&25.25&0.37\\
21:08:48.88&-51:09:58.27&0.234&25.50&1.49&&21:08:51.40&-51:10:38.67&0.864&25.90&0.52\\
21:08:48.46&-51:09:45.11&0.244&25.59&0.51&&21:08:51.27&-51:08:57.01&0.884&23.95&1.56\\
21:08:48.39&-51:09:46.36&0.250&23.92&1.08&&21:08:45.25&-51:10:20.17&0.907&25.32&0.31\\
21:08:49.60&-51:10:04.35&0.269&26.00&0.50&&21:08:53.34&-51:09:04.04&0.912&26.30&0.67\\
21:08:49.67&-51:09:32.71&0.269&25.70&0.50&&21:08:44.72&-51:10:12.36&0.914&22.54&0.59\\
21:08:48.30&-51:09:44.43&0.270&24.17&0.36&&21:08:46.21&-51:09:06.23&0.920&25.77&0.49\\
21:08:51.71&-51:09:52.70&0.282&25.53&0.27&&21:08:44.95&-51:10:18.04&0.927&20.80&0.71\\
21:08:48.41&-51:09:57.73&0.288&24.73&0.36&&21:08:47.23&-51:08:58.48&0.939&24.63&1.43\\
21:08:49.48&-51:10:05.34&0.289&24.04&0.17&&21:08:55.76&-51:10:04.42&0.945&24.19&0.35\\
21:08:51.85&-51:09:46.59&0.297&25.15&0.28&&21:08:54.01&-51:10:31.06&0.950&20.43&0.53\\
21:08:51.62&-51:09:58.41&0.306&25.11&0.30&&21:08:44.83&-51:09:16.29&0.969&23.59&0.70\\
21:08:50.31&-51:09:30.40&0.308&23.70&1.60&&21:08:54.60&-51:09:08.32&0.989&26.07&0.39\\
21:08:50.67&-51:10:06.09&0.312&25.57&0.31&&21:08:52.91&-51:10:41.16&0.990&25.63&0.28\\
21:08:51.88&-51:09:42.67&0.316&24.80&0.30&&21:08:55.70&-51:09:21.85&1.003&24.48&0.37\\
21:08:48.88&-51:10:04.68&0.318&25.27&0.42&&21:08:56.36&-51:09:43.54&1.005&25.58&0.92\\
21:08:50.68&-51:09:30.35&0.324&25.26&0.34&&21:08:49.09&-51:08:47.63&1.025&25.96&0.77\\
21:08:51.56&-51:09:35.91&0.328&25.90&0.43&&21:08:56.25&-51:10:08.85&1.040&24.63&0.31\\
21:08:49.14&-51:09:30.18&0.333&24.97&0.44&&21:08:44.24&-51:10:20.47&1.043&23.33&1.35\\
21:08:47.66&-51:09:47.41&0.363&26.41&0.39&&21:08:43.38&-51:10:01.49&1.054&25.58&0.22\\
21:08:51.13&-51:09:29.50&0.367&22.09&0.63&&21:08:56.65&-51:09:59.53&1.063&26.09&0.81\\
21:08:50.39&-51:10:10.67&0.374&24.61&1.44&&21:08:48.50&-51:08:46.22&1.065&25.86&1.00\\
21:08:49.91&-51:10:11.50&0.382&23.73&1.23&&21:08:43.26&-51:10:00.83&1.071&26.09&0.43\\
21:08:52.32&-51:09:42.66&0.382&26.13&0.82&&21:08:55.74&-51:10:24.13&1.082&24.03&0.35\\
21:08:47.88&-51:10:01.51&0.391&23.16&0.44&&21:08:47.64&-51:10:49.72&1.082&23.07&0.54\\
21:08:52.53&-51:09:53.25&0.409&24.85&0.39&&21:08:49.07&-51:08:43.27&1.098&25.99&-0.18\\
21:08:52.60&-51:09:57.71&0.440&26.42&0.55&&21:08:50.56&-51:08:42.41&1.107&24.78&1.10\\
21:08:52.15&-51:09:31.75&0.443&24.86&0.23&&21:08:43.01&-51:10:02.77&1.117&26.45&0.16\\
21:08:50.66&-51:09:21.20&0.469&25.66&0.47&&21:08:44.16&-51:09:09.61&1.119&24.78&1.60\\
21:08:47.68&-51:09:25.65&0.525&22.35&0.63&&21:08:46.02&-51:10:45.91&1.138&23.16&1.48\\
21:08:46.90&-51:09:33.69&0.542&26.36&0.51&&21:08:55.17&-51:10:36.31&1.139&23.78&0.93\\
21:08:47.50&-51:10:12.08&0.551&22.93&1.43&&21:08:56.24&-51:09:14.14&1.140&25.06&0.39\\
21:08:46.41&-51:09:50.38&0.558&26.21&0.60&&21:08:48.42&-51:10:55.57&1.142&26.18&0.72\\
21:08:47.03&-51:09:29.66&0.558&26.19&0.68&&21:08:42.91&-51:09:29.48&1.152&25.07&0.40\\
21:08:46.65&-51:09:35.71&0.562&26.35&0.54&&21:08:45.39&-51:10:42.69&1.152&24.78&1.82\\
21:08:52.89&-51:10:08.28&0.564&22.95&0.58&&21:08:53.20&-51:08:45.94&1.161&25.63&0.50\\
21:08:52.23&-51:09:22.30&0.564&25.39&0.54&&21:08:46.98&-51:08:44.46&1.167&23.50&1.46\\
21:08:51.70&-51:09:17.53&0.585&25.84&0.61&&21:08:43.92&-51:10:29.53&1.168&26.26&0.44\\
21:08:53.72&-51:09:50.90&0.590&26.29&0.65&&21:08:50.94&-51:08:38.12&1.184&24.01&1.50\\
21:08:50.92&-51:09:14.21&0.592&25.23&1.39&&21:08:57.20&-51:09:20.16&1.229&24.74&0.43\\
21:08:53.75&-51:09:49.88&0.593&26.49&0.42&&21:08:45.21&-51:08:49.99&1.230&26.37&0.34\\
21:08:53.24&-51:09:30.70&0.593&24.77&1.69&&21:08:42.36&-51:10:06.62&1.230&26.48&0.35\\
21:08:46.27&-51:09:40.95&0.593&25.36&0.38&&21:08:45.70&-51:08:46.41&1.234&24.63&1.50\\
21:08:46.16&-51:09:53.32&0.603&25.63&0.88&&21:08:51.29&-51:08:35.17&1.241&25.50&1.24\\
21:08:47.36&-51:10:15.60&0.608&25.87&0.31&&21:08:43.39&-51:09:05.60&1.256&23.35&1.48
 \label{tab:GMOSdata}
 
\end{deluxetable*}

\subsection{Age and Metallicity}

\begin{figure}
\begin{center}
\includegraphics[scale=0.45]{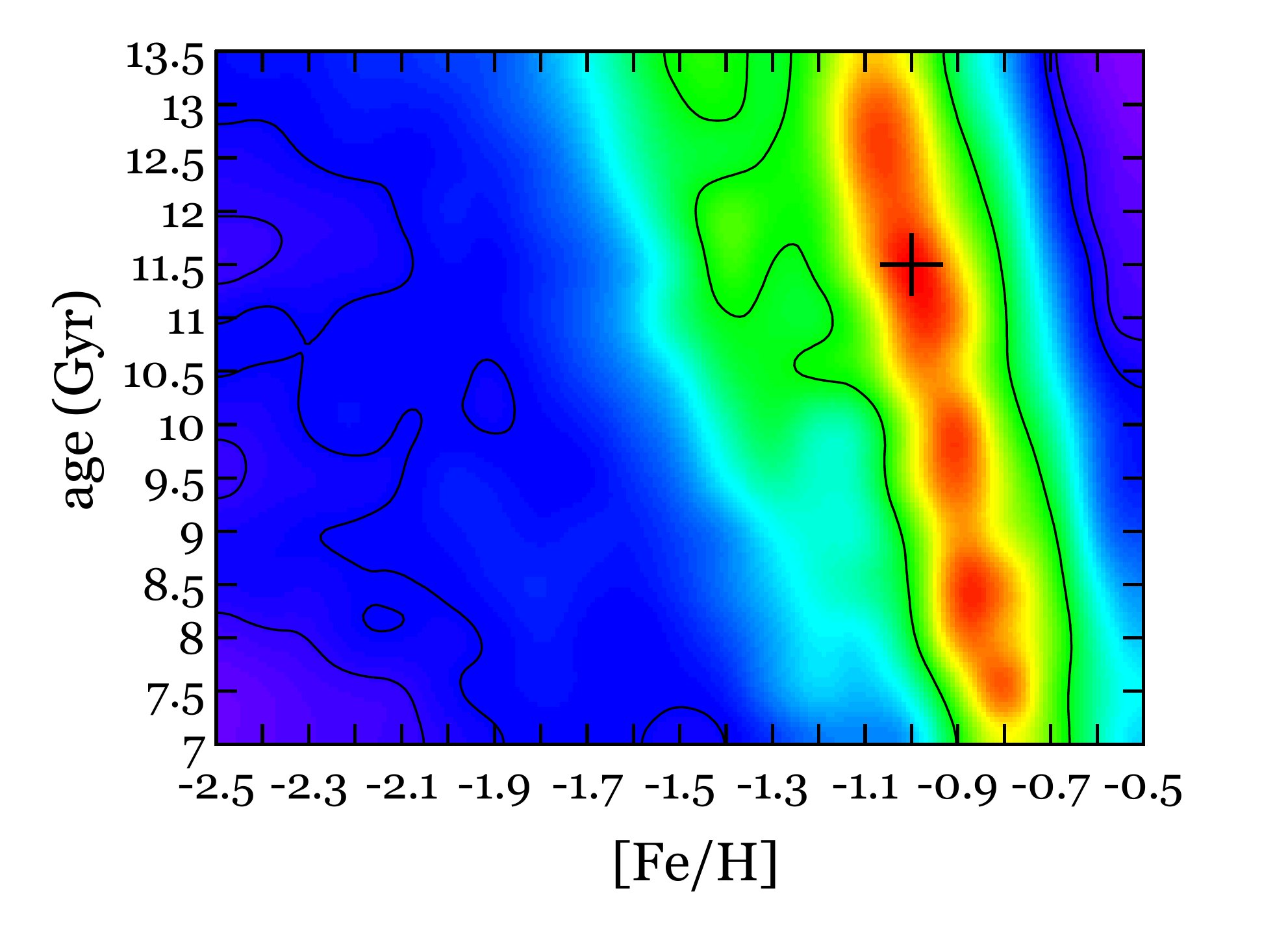}
\end{center}
\caption{Smoothed maximum likelihood density map in age-metallicity space for all stars within a radius 
of $1.4\arcmin$ around Kim\,2. Contour lines show the 68\%, 95\%, and 99\% confidence levels. 
The diagonal flow of the contour lines reflect the age-metallicity degeneracy inherent to such an isochrone
fitting procedure. The 1D marginalized parameters around the best fit are:  age $=11.5^{+2.0}_{-3.5}$\,Gyr, 
[Fe/H]$=-1.0^{+0.18}_{-0.21}$\,dex, 
[$\alpha$/Fe]=$+0.2$\,dex,
$m-M=20.10\pm0.10$\,mag.}
\label{fig:MLplot}
\end{figure}

We estimate the age, metallicity, alpha element to iron abundance, and distance of Kim\,2 using the maximum likelihood method described 
in~\cite{Frayn2002}, \cite{Fadely2011}, and \cite{Kim1}. For the analysis we use all stars within a radius of $1\farcm4$ around Kim\,2,
the inner circle in the left panel of Figure~\ref{fig:CMD}.  We calculate the maximum likelihood values $\mathcal{L}_i$, as 
defined by the Equations\,1 and 2 in Fadely et al. (2011), over a grid of Dartmouth model isochrones \citep{Dartmouth}, where
$i$ symbolises the grid points in the multi-dimensional parameter space that covers the age range from 7.0--13.5\,Gyr, a metallicity 
range $-2.5\leq$ [Fe/H] $\leq-0.5$\,dex, $-0.2\leq [\alpha/$Fe$]\leq +0.4$ and a distance range 
$19.5<(m-M)<20.5$.  Grid steps are 0.5\,Gyr, 0.1\,dex, 0.2\,dex, and 0.05\,mag, respectively.  

In Figure~\ref{fig:MLplot}, we present the matrix 
of likelihood values for the sample described above after interpolation and smoothing over two grid points. Depending on the 
weight given to the two MSTO stars (labelled 1 and 2 in Figure~\ref{fig:CMD}), we find two slightly different isochrones that fit the data best
(center panel of Figure~\ref{fig:CMD}).
If the two stars are given the weights based on their photometric uncertainties, the best-fitting isochrone has an age of 
11.5 Gyr and [Fe/H] $= -1.0$\,dex, $[\alpha/$Fe$]=+0.2$, with a heliocentric distance of 104.7\,kpc ($(m-M)=20.10$), 
the solid blue line in Figure~\ref{fig:MLplot}. However, if we give extra weight to these stars because they are close to 
the cluster center we derive a younger age of 8.0 Gyr, [Fe/H] $= -0.9$\,dex, $[\alpha/$Fe$]=+0.4$, 
and a distance of 98\,kpc ($(m-M)=19.96$). The probability that goes with this second solution is 0.9\% lower than the first solution.
In the following we will adopt the parameters of the first solution. In particular, we use a heliocentric distance of 
105\,kpc for Kim\,2 in the calculation of the physical size and absolute magnitude (Section 4.3).
The 68\%, 95\%, and 99\% confidence contours are 
overplotted in Figure \ref{fig:MLplot}.  

The marginalized uncertainties about this most probable location correspond to an age of 
$11.5^{+2.0}_{-3.5}$\,Gyr, a metallicity of [Fe/H]$=-1.0^{+0.18}_{-0.21}$\,dex, and a distance modulus of 
$\rm (m-M)_0=20.10\pm0.10$\,mag ($D_\odot = 104.7\pm4.1$ kpc). For the 2nd solution we get: 
$8.0^{+3.5}_{-2.0}$\,Gyr, [Fe/H]$=-0.9^{+0.16}_{-0.19}$\,dex, 
and $\rm (m-M)_0=19.96\pm0.09$\,mag ($D_\odot = 98.2\pm4.2$ kpc), respectively.

In the discussion about the age and metallicity of Kim\,2 it is important to note that a significant fraction of unresolved MS-MS binaries are 
common among low-mass star clusters.
Some low-luminosity clusters in the \cite{Milone2012} sample like E3 have 50\% or more binaries (see their Fig.\,B.4). As we will show in Section 4.3,
Kim 2 is among the lowest luminosity (hence lowest mass) star clusters known. If Kim\,2 follows the anti-correlation between mass and binary fraction then it would host a large population of binaries. Unfortunately, our photometry does not allow us to distinguish binaries from single 
MS stars in the CMD, but we know that binaries are located on the red side of the MS. This means that, due to the present of many binaries, the MS 
we observe would be redder than the MS of single stars. Furthermore, binaries of turn-off stars can be located above the turn-off similar to 
the observed stars we coloured in red in the CMD. Hence, if Kim\,2 has a high binary fraction, we are likely to over-estimate the metallicity
by about 0.2-0.4\,dex. A detailed study of the binary fraction in Kim\,2 shall be the focus of an upcoming study.

Both of our two age/metallicity solutions for Kim\,2 are comparable to the observed properties of globular clusters seen in the Galactic halo. A large ensemble of these objects have been studied in detail with the Hubble Space Telescope -- see, for example,  Figure 13 in \cite{Marin-Franch2009}. At [Fe/H]$\sim -1.0$  and [$\alpha$/Fe] $\sim +0.2$ ([M/H]$\approx -0.86$) we find clusters as young as 0.8 times the age of the oldest systems in the Milky Way; an age of 11.5 Gyr would thus be consistent with with many other Galactic GCs (albeit lying at smaller Galactocentric radii). Moving to the second solution, at  [Fe/H]$\sim-0.9$ or $-0.8$ and  [$\alpha$/Fe] $\sim +0.4$ ([M/H]$\approx -0.56$), there are clearly somewhat younger clusters than at [M/H]$\sim-0.86$, but an age of 8.0 Gyr would be certainly on the younger envelope of the observed distribution (this would correspond to a relative age of about 0.6 or so). Note that by this metallicity the age-metallicity relation for Galactic GCs has clearly bifurcated. Clusters on the upper locus are thought to be accreted objects, and Kim 2 would clearly be part of that ensemble. The other point worth mentioning is that [$\alpha$/Fe] $\sim +0.4$ at this age and metallicity would be unusually higher than many comparable Galactic GCs. For example, objects with similar ages and metallicities such as Terzan\,7 and Pal\,12, have [$\alpha$/Fe] $\sim0.0$. Higher [$\alpha$/Fe]  would suggest this object came from a relatively massive parent galaxy (perhaps LMC mass or so) where the chemical enrichment proceeded quite quickly, i.e, where the ``knee'' in the [$\alpha$/Fe]  vs. [Fe/H] plot is at comparatively high [Fe/H]. For the older solution [$\alpha$/Fe] $\sim+0.2$ and [Fe/H] $\sim -1$ would seem to be normal compared to other similar Galactic GCs.

\subsection{Size, Ellipticity, and Luminosity}

\begin{figure*}
\begin{center}
\includegraphics[scale=0.85]{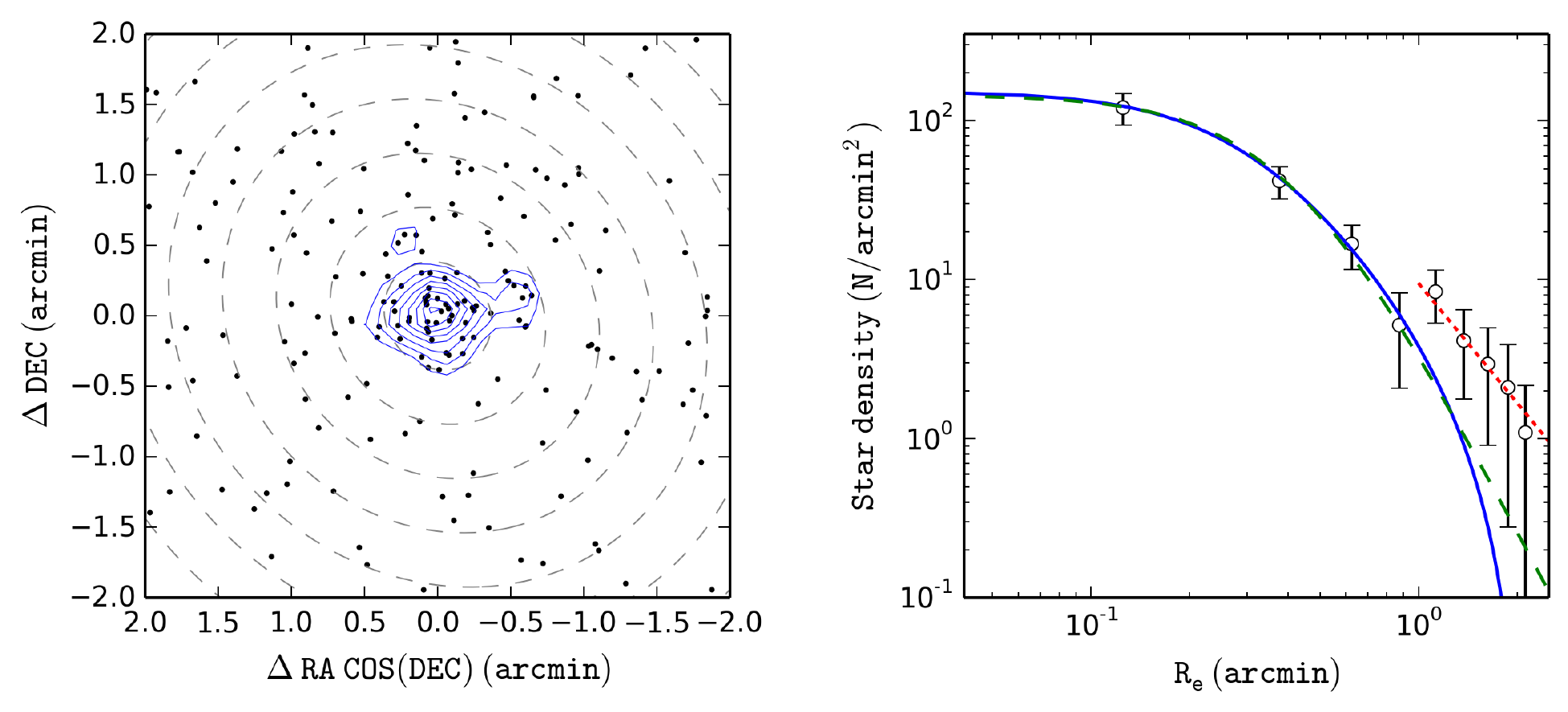}
\end{center}
\caption{\emph{Left panel}: RA-DEC distribution of GMOS stars  centred on Kim\,2. The dashed ellipses indicate $0.4\,\arcmin$ steps in the elliptical radius of ellipcity $\epsilon=0.12$ and position angle $\theta=35\,\deg$. The solid contours mark the local star density that is $3-10\sigma$  above the background density. \emph{Right panel}: Radial stellar density profile based on the stellar distribution in the left panel. Open circles represent the star density after subtracting the background contribution. The solid line, dashed line and dotted line show a King profile with a core radius $r_{c}=0\farcm28$ and tidal radius $r_{t}=2\farcm1$, a Plummer profile with a half-light radius $r_{h}=0\farcm42$, and a power law profile with a slope $\gamma=-2.5$ for the extra tidal stars, respectively.} \label{fig:RadialProfile}
\end{figure*}

The left panel of Figure~\ref{fig:RadialProfile} shows the sky distribution of all GMOS stars in the magnitude range $23.5<r<26.5$ and $g<27.0$ in a $0.7^\circ\times0.7^\circ$ window centred on Kim\,2. The solid contours at the centre correspond to $3-10\sigma$ levels above the background density. We derive an ellipticity $\epsilon=0.12$ and the position angle $\theta=35\deg$ using the fit\_bivariate\_normal function of the astroML package~\citep{astroML}. The right panel shows the associated radial profile of the stellar number density, where $R_{e}$ is the elliptical radius. To estimate the background of field stars, we subtracted from the catalogue the stars consistent with the isochrone and counted the remaining stars in the same color-magnitude range, which results in $6.4$ stars per sqr arcmin. The error bars were derived based on Poisson statistics. The best-fit King profile based on the innermost four data points gives a core radius of $0\farcm28\pm0.02$ or $r_{c}=8.5\pm0.6$\,pc, and a tidal radius of $2\farcm10\pm0.02$ or $r_{t}=64.0\pm0.6$\,pc adopting the distance modulus of 20.1\,mag. We note that the observed radial profile exceeds the King model at radii $R_{e}>1\farcm0$. Such a departure from the King model at radii considerably less than $r_{t}$ has been already reported for many other globular clusters and identified as extra-tidal stars that follow a power law density profile~\citep[e.g.][]{Grillmair1995,Carraro2007,Carraro2009}. We also estimated a half light radius by means of the best-fitting Plummer profile, which yields $0.42\pm0.02$\,arcmin or $r_{h}=12.8\pm0.6$\,pc (dashed line). Above the King and Plummer profiles, we outline the extra-tidal stars using a power law profile with slope $\gamma=-2.5$ (dotted line). 

We further derived the total magnitude of Kim\,2 as follows. We selected Kim\,2 stars by means of a photometric filter based on the best-fitting Dartmouth isochrone and taking into account the uncertainties of our photometry. We then built the observed luminosity function (LF) of Kim\,2 by counting the selected stars within $3r_{h}$ as a function of magnitude from the saturation level $r_0=19.5$ to the 50\% completeness limit $r_0=26.5$. The observed LF was corrected for  incompleteness.  We then adopted a normalized theoretical LF based on Dartmouth model~\citep{Dartmouth} and scaled it to the observed LF, for which we used two scale factors: (1) the ratio of the integrated number density of the observed LF to the probability density of a normalised theoretical LF between the saturation and  the 50\% completeness limits  and (2) the ratio of the integrated flux of the observed LF to that of the theoretical LF in the same magnitude range. We obtained the total magnitude by integrating the scaled theoretical LF inclusive of the missing flux at $r$ magnitude fainter than 50\% completeness limit. Method (1) yields $M_{r}=-1.74$\,mag and method (2) $M_{r}=-1.73$\,mag. The total V-band luminosity is therefore $M_{V}=-1.47$\,mag ($V-r=0.264$, adapted from Dartmouth model for a 11.5 Gyr, [Fe/H] = - 1.0 stellar population). Since we included all stars consistent with the isochrone, the calculated value should be considered an upper limit of the true total V-band luminosity and an exclusion of a single RGB star can change it still by $\sim0.5$\,mag. This result suggests that Kim\,2 is among the faintest known Milky Way globular clusters, with a comparable luminosity to Balbinot\,1~\cite[$M_{V}\sim-1.21$;][]{Balbinot2013}, Koposov\,1 ($M_{V}\sim-1.35$), AM\,4 ($M_{V}\sim-1.8$). The only MW star clusters with even lower luminosities are Mu\~noz\,1~\cite[$M_{V}\sim-0.4$;][]{Munoz2012}, Koposov\,2~\cite[$M_{V}\sim-0.4$;][and 2010 edition]{Harris1996}, Segue\,3~\cite[$M_{V}\sim0.0$;][]{Fadely2011}, and Kim\,1~\cite[$M_{V}\sim0.3$;][]{Kim1}. 
All derived parameters for Kim\,2 are summarised in Table~\ref{tab:Parameters}. 

\begin{deluxetable}{lrl}
\tablewidth{0pt}
\tablecaption{Properties of Kim\,2}
\tablehead{
\colhead{Parameter} & 
\colhead{Value} &
\colhead{Unit}}
\startdata
$\alpha_{J2000}$ & 21:08:49.97 & h:m:s\\
$\delta_{J2000}$ & -51:09:48.60 & $^\circ:\arcmin:\arcsec$\\
$l$ & $347.159$ & deg\\
$b$ & $-42.074$ & deg\\
$(m-M)_0$ & $20.10\pm0.10$ & mag \\       
$D_\odot$ & $104.7\pm4.1$ & kpc \\          
$D_{gc}$ & $99.4\pm3.9$ & kpc \\          
{[Fe/H]} & $-1.0^{+0.18}_{-0.21}$ & dex \\
{[$\alpha$/Fe]} & $+0.2$ & dex \\
Age & $11.5^{+2.0}_{-3.5}$ & Gyr \\
$r_{h}$(Plummer) & $12.8\pm0.6$ \tablenotemark{a} & pc \\
$r_{c}$(King) & $8.5\pm0.6$ \tablenotemark{a} & pc \\
$r_{t}$(King) & $64.0\pm0.6$ \tablenotemark{a} & pc \\
$\epsilon$ & $0.12\pm0.10$ & \\
$\theta$ & $35\pm5$ & deg \\
$M_{tot,V}$ & $-1.5\pm0.5$ & mag 
\enddata
\tablenotetext{a}{ Adopting a distance of 104.7\,kpc}
\label{tab:Parameters}
\end{deluxetable}

\section{Discussion and Conclusion}

We report the discovery of a star cluster, Kim\,2, in the outer halo of the MW. This object was first detected in our DECam blind field survey data and confirmed by GMOS follow-up observation. We found the cluster distant ($\sim100$\,kpc), faint ($M_{V}\sim-1.4$), younger than the oldest GCs ($<11.5$Gyr) and more metal-rich
 ([Fe/H]$\sim-1.0$) than any outer halo GC. Its physical size ($r_{h}\sim12.8$\,pc) is comparable to that of the other outer halo GCs but with an order of magnitude difference in terms of luminosity~\cite[see Figure 8 and 9 in][]{Mackey2005}. 

\subsection{Evidence of Mass Loss}
\begin{figure}
\begin{center}
\includegraphics[scale=0.65]{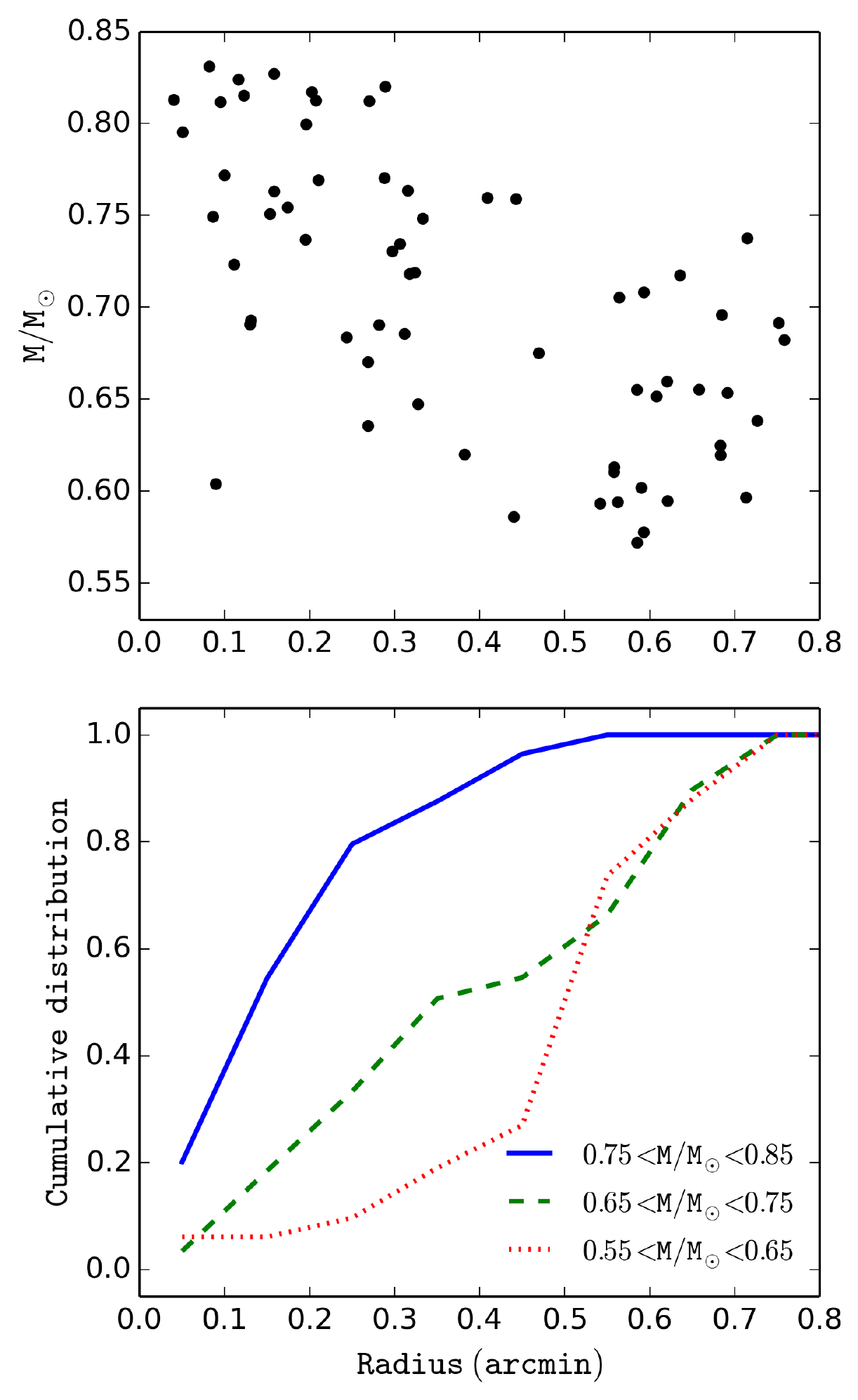}
\end{center}
\caption{ Upper panel: stellar mass of all Kim\,2 main sequence stars within $2r_{h}$($\sim0\farcm8$) as a function of the distance from the cluster center. Lower panel: Cumulative distribution function of Kim\,2 main sequence stars for three different mass intervals.\label{fig:MassSeg}}
\end{figure}

Low luminosity globular clusters are expected to be in a mass segregation state as the relaxation time of the clusters is significantly 
shorter than their respective ages. To estimate the half-mass two-body relaxation time $t_{rh}$ of Kim\,2 we used the following equation~\citep{Spitzer1971}, 

\begin{equation}
t_{rh}=\frac{8.9\times10^{5}M^{1/2}R_{h}^{3/2}}{\bar{m}\log_{10}(0.4M/\bar{m})}
\end{equation}

where $M$ is the mass of a cluster, $R_{h}$ is the radius containing half mass of the cluster, $\bar{m}$ is the average mass of the members. Here, we estimated the cluster mass $M\sim600M_{\odot}$ and the average stellar mass $\bar{m}\sim0.3M_{\odot}$ using an initial mass function by \cite{Chabrier2001} and an isochrone by \cite{Parsec}. We used the half-light radius of $12.8$\,pc for $R_{h}$. Using these numbers gives the relaxation time $t_{rh}\sim1.1$\,Gyr, which is significantly short relative to the estimated age of the cluster($\sim11.5$\,Gyr). This result suggests that Kim\,2 should have had sufficient time to undergo dynamical mass segregation. To investigate this possibility
we show in the upper panel of Figure~\ref{fig:MassSeg} the mass of the stars consistent with the main sequence of Kim\,2 in the magnitude range $23.5<r_{0}<26.5$ as a function of the distance from the center of the cluster. The stellar mass systematically decreases over the radius. The lower panel shows the normalized cumulative distribution function for three different mass intervals ($0.55<M/M_{\odot}<0.65$, $0.65<M/M_{\odot}<0.75$, and $0.75<M/M_{\odot}<0.85$), corrected for incompleteness. The plots clearly show that more massive MS stars preferentially populate the inner part of the cluster. 

\begin{figure}
\begin{center}
\includegraphics[scale=0.8]{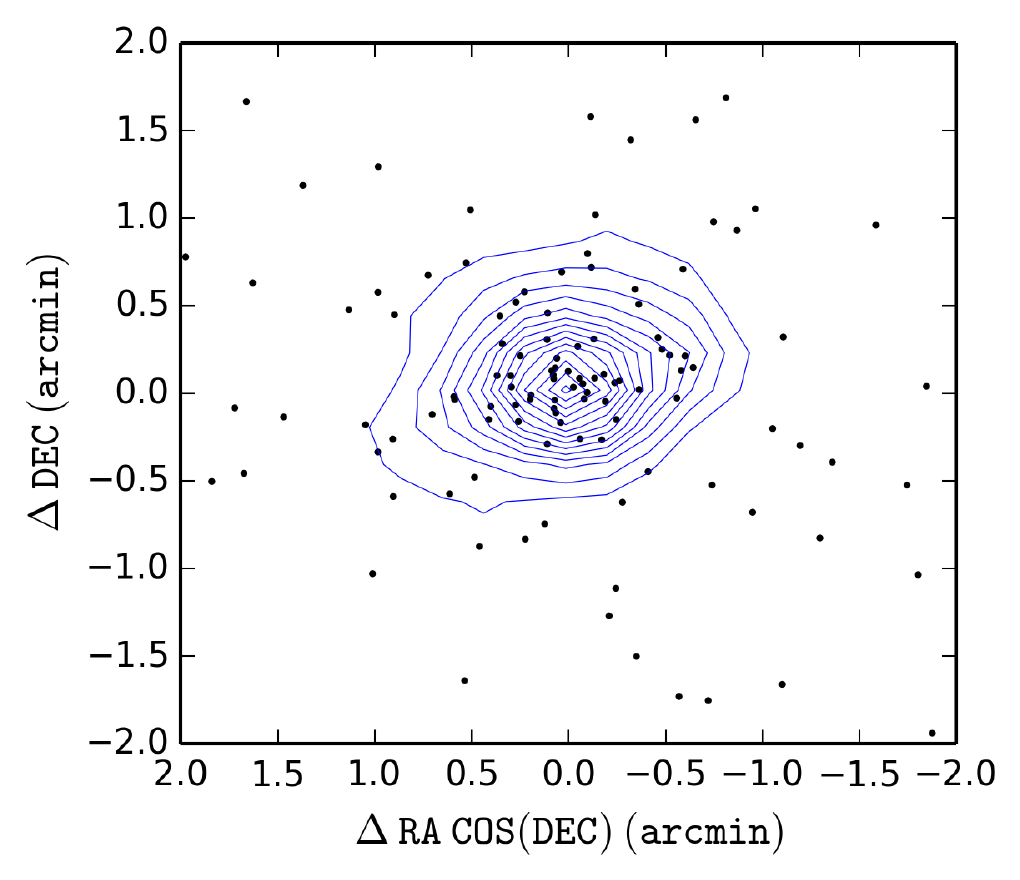}
\end{center}
\caption{Distribution of stars consistent with Kim\,2's main sequence . Contours mark $1.0-7.5\sigma$ levels above the background, with a step size of $0.5\sigma$.\label{fig:TidtalStructure}}
\end{figure}

Figure~\ref{fig:TidtalStructure} shows the distribution of the potential MS stars and contours of $1-6\sigma$ levels above the background. Although no tail structure of the extra tidal stars at radius $>1\farcm0$ is noticeable, the outer contours are slightly more elliptical in a rather consistent orientation ($\theta\sim105\deg$). Similar changes
of the orientation angle are observed in the cores of other GCs undergoing tidal disruption~\citep[e.g. Pal1, Pal\,5 and Pal\,14 in][]{Niederste-Ostholt2010,Odenkirchen2001,Sollima2011}. Considering the low concentration and luminosity of Kim\,2, these stars are likely to be loosely bound around the center.

These results suggest that Kim\,2 must have experienced substantial mass loss by relaxation and tidal stripping in order to have reached its current physical and dynamic state. With the consistency between the two-body relaxation time and the observed mass segregation, it seems unlikely  that Kim\,2 contains any significant amount of dark-matter, since otherwise the half-mass radius and the total mass of the system would be much greater than observed leading to a relaxation time comparable to or even exceeding the Hubble time. Accordingly, dynamical mass segregation would be hardly observed in  a dark-matter dominated stellar system.

\subsection{Is Kim\,2 associated with a MW satellite galaxy or Stream?}
Kim\,2 has an unusually low luminosity when compared with the other known outer halo GCs (Table~\ref{tab:outerGCs}). 
The significant luminosity difference of at least $3$\,mag and evidence of extra-tidal stars (Section 4.3) strongly 
suggests we are seeing an outer halo GC that experienced mass loss due to the MW tidal field, similar to Pal\,14~\citep{Sollima2011}. Kim\,2 and Pal\,14 share low star density and evidence of tidal interaction. As suggested by \cite{Sollima2011}, an outer halo GC with such low star density is likely to follow an orbit confined to the outer region of the Galactic halo, and/or to have formed in a dwarf galaxy that was later accreted into the Galactic halo. As a consequence, the cluster could have experienced minor tidal disruption and survived until the present epoch.

Kim 2 also shares properties with AM4 in terms of metallicity, age and luminosity~\citep{Carraro2009}. 
AM4 is considered to be associated with the Sagittarius (Sgr) dwarf galaxy. Kim\,2 is approximately 25$^\circ$ away from the 
orbit of the Sgr tidal stream, and the \cite{Law2010} model of the Sgr dwarf galaxy embedded in a triaxial MW halo 
has only a single Sgr Stream particle at a heliocentric distance of 79\,kpc within 0.5 sqr deg of Kim\,2.  This very low particle
density and the large line-of-sight distance difference of 26\,kpc makes it highly unlikely that Kim\,2 originates from the Sgr galaxy. 

However, there is still the possibility that Kim\,2 is not a genuine MW globular cluster, but was formerly associated with another dwarf 
galaxy, which deposited it into the Galactic halo. In that context, we note that Kim\,2 is relatively close to the vast polar structure (VPOS), a thin 
(20\,kpc) plane perpendicular to the MW disk defined by the 11 brightest Milky Way satellite 
galaxies~\citep{Kroupa2005,Metz2007,Metz2009,Kroupa2010,Keller2012,Pawlowski2012}. In the region of Kim\,2 VPOS is defined by the Small and Large Magellanic clouds, 
and Carina.Globular clusters and stellar and gaseous streams appear to preferentially align with the VPOS too~\citep{Forbes2009,Pawlowski2012}. 
The origin of that plane is still a matter of debate. It could be the result of a major galaxy collision that left debris in form of tidal dwarfs and star clusters along the orbit~\citep{Pawlowski2013}. A more detailed analysis of this matter is beyond the scope of this paper due 
to the small field of view of GMOS and the shallowness of our more extended DECam photometry.

\acknowledgements{The authors like to thank Tammy Roderick, Kathy Vivas and David James for their assistance during the DECam observing run. We also thank the referee for the helpful comments and suggestions, which contributed to improving the quality of the publication. We acknowledge the support of the Australian Research Council (ARC) through Discovery projects DP120100475, DP150100862, DP1093431 and financial support from the Go8/Germany Joint Research Co-operation Scheme. 

Based on observations obtained at the Gemini Observatory, processed using the Gemini IRAF package and gemini\_python.
Gemini Observatory is operated by the Association of Universities for Research in Astronomy, Inc., under a cooperative agreement with 
the NSF on behalf of the Gemini partnership: the National Science Foundation (United States), the National Research Council (Canada), 
CONICYT (Chile), the Australian Research Council (Australia), Minist\'{e}rio da Ci\^{e}ncia, Tecnologia e Inova\c{c}\~{a}o 
(Brazil) and Ministerio de Ciencia, Tecnolog\'{i}a e Innovaci\'{o}n Productiva (Argentina).

This project used data obtained with the Dark Energy Camera (DECam), which was constructed by the Dark Energy Survey (DES) collaborating institutions: Argonne National Lab, University of California Santa Cruz, University of Cambridge, Centro de Investigaciones Energeticas, Medioambientales y Tecnologicas-Madrid, University of Chicago, University College London, DES-Brazil consortium, University of Edinburgh, ETH-Zurich, University of Illinois at Urbana-Champaign, Institut de Ciencies de l'Espai, Institut de Fisica d'Altes Energies, Lawrence Berkeley National Lab, Ludwig-Maximilians Universitat, University of Michigan, National Optical Astronomy Observatory, University of Nottingham, Ohio State University, University of Pennsylvania, University of Portsmouth, SLAC National Lab, Stanford University, University of Sussex, and Texas A$\&$M University. Funding for DES, including DECam, has been provided by the U.S. Department of Energy, National Science Foundation, Ministry of Education and Science (Spain), Science and Technology Facilities Council (UK), Higher Education Funding Council (England), National Center for Supercomputing Applications, Kavli Institute for Cosmological Physics, Financiadora de Estudos e Projetos, Funda\c{c}\~ao Carlos Chagas Filho de Amparo a Pesquisa, Conselho Nacional de Desenvolvimento Cient\'i­fico e Tecnol\'ogico and the Ministério da Ci\^encia e Tecnologia (Brazil), the German Research Foundation-sponsored cluster of excellence "Origin and Structure of the Universe" and the DES collaborating institutions.

%\bibliographystyle{apj}
%\bibliography{paper}

\end{document}